\documentclass{article}
\usepackage{ijcai22}

\usepackage{times}

\usepackage{soul}
\usepackage{url}
\usepackage[hidelinks]{hyperref}
\usepackage[utf8]{inputenc}
\usepackage[small]{caption}
\usepackage{graphicx}
\usepackage{amsmath}
\usepackage{booktabs}
\usepackage{color}
\usepackage{amsthm}
\usepackage{amsmath}
\usepackage{amssymb}
\theoremstyle{definition}

\usepackage{amsthm}

\usepackage{mathtools}
\usepackage[T1]{fontenc}
\usepackage{comment}
\usepackage{bm}
\usepackage{subcaption}
\usepackage{booktabs}
\usepackage{xcolor}
\usepackage{threeparttable}
\usepackage[ruled,linesnumbered]{algorithm2e}
\newcommand*{\defeq}{\stackrel{\text{def}}{=}}

\let\oldnl\nl
\newcommand{\nonl}{\renewcommand{\nl}{\let\nl\oldnl}}
\SetKwInput{KwData}{Input}
\SetKwInput{KwResult}{Output}
\DontPrintSemicolon

\usepackage{tikz}

\usepackage{latexsym} 

\newtheorem{proposition}{Proposition}
\newtheorem{lemma}{Lemma}

\title{\textsc{Chem}: Efficient Secure Aggregation with Cached Homomorphic Encryption in Federated Machine Learning Systems}
\author{Dongfang Zhao
\emails 
University of Nevada, Reno
}

\begin{document}

\maketitle

\begin{abstract}
Although homomorphic encryption can be incorporated into neural network layers for securing machine learning tasks,
such as confidential inference over encrypted data samples and encrypted local models in federated learning,
the computational overhead has been an Achilles heel.
This paper proposes a caching protocol, namely \textsc{Chem}, 
such that tensor ciphertexts can be constructed from a pool of cached radixes rather than carrying out expensive encryption operations.
From a theoretical perspective, we demonstrate that \textsc{Chem} is semantically secure and can be parameterized with straightforward analysis under practical assumptions.
Experimental results on three popular public data sets show that adopting \textsc{Chem} only incurs sub-second overhead and yet reduces the encryption cost by 48\%--89\% for encoding input data samples in confidential inference and 67\%--87\% for encoding local models in federated learning, respectively.
\end{abstract}

\section{Introduction}

\subsection{Background and Motivation}

While increasingly more applications have outsourced their computing infrastructure to cloud computing vendors,
this paradigm inevitably causes serious concerns over data security.
Among others, machine learning tasks have been striving to achieve both high accuracy and high confidentiality on remote computing resources.
To that end, adopting encryption in machine learning tasks has drawn tremendous research interest;
example tasks include applying a trained model to encrypted test data (i.e., encrypted inference) and aggregating (encrypted) local models in federated learning~\cite{bmcmahan_aistats17}.
One plausible approach is to adopt a specific type of encryption scheme, 
namely fully homomorphic encryption (FHE)~\cite{cgentry_stoc09},
which allows arbitrary operations built upon the basic arithmetic additive and multiplicative operations directly on the ciphertexts.
Some notable systems for employing FHE in machine learning include CryptoNets~\cite{ndowlin_icml16}, FHE–DiNN~\cite{fbour_crypto18}, and GAZELLE~\cite{cjuve_sec18}.

Although FHE can be incorporated into neural network layers in machine learning,
the computational overhead has been an Achilles' heel of outsourced machine learning systems:
FHE involves a sophisticated procedure involving hard mathematical problems that are computation-intensive.
To mitigate the performance overhead of FHE,
Glyph~\cite{qlou_nips20} proposed to switch between FHE schemes to reduce the training cost of deep neural networks (DNNs).
Similarly, various works~\cite{qlou_icml21,qlou_nips20a,qlou_nips20b,qliu_nips19,rdath_pldi19,sriazi_sec19} aimed to reduce the inference latency.

While the existing works have focused on the network performance over encrypted data,
the data encoding itself---i.e., encrypting the tensor plaintext into ciphertext---is treated as a black box with little optimization.
As a result, the encoding procedure could unexpectedly become the bottleneck of the entire machine-learning system,
as we will show later in~Figure~\ref{fig:performance} of this paper
(e.g., encoding 100 MNIST~\cite{mnist} images takes unacceptable 549 seconds using the latest FHE library SEAL~\cite{seal}).
Caching existing HE ciphertexts is also being actively researched in the database community~\cite{otawose_sigmod23},
especially in outsourced databases.
Other techniques orthogonal than homomorphic encryption for privacy-preserving computation include secret sharing~\cite{cdwork_tcc06} and multiparty computation~\cite{ayao_focs82},
which are not discussed in this paper.

\subsection{Contributions}

This work aims to reduce the latency of tensor encryption in machine learning tasks without compromising the security levels.
This work makes the following contributions:
\begin{itemize}

\item We propose a new caching protocol such that tensor ciphertexts can be constructed from a pool of cached radixes rather than carrying out the expensive encryption operations.
The insight of the caching protocol lies in the fact that FHE ciphertexts can be expanded with an arbitrary radix base along with the homomorphic additive operation.
To achieve low latency, 
we cache only a logarithmic number of radixes.

\item We prove the security property of the proposed protocol for caching tensors in machine learning tasks.
In particular, we articulate the threat model,
i.e., what types of attacks and computational resources we assume an adversary may have.
Then, we theoretically demonstrate the IND-CPA security of the proposed caching protocol under practical assumptions.

\item We conduct a thorough analysis of the parameterization of the proposed caching protocol regarding the choice of radix.
We show that although the radix can be set arbitrarily from a theoretical point of view,
the optimal radix turns out to be binary for many machine-learning tasks.
\end{itemize}

We have implemented the proposed caching protocol, namely \textsc{Chem}, 
with OpenMined TenSEAL~\cite{tenseal2021} library that is built upon Microsoft SEAL~\cite{seal}.
We have evaluated \textsc{Chem} with two baseline FHE schemes (CKKS~\cite{ckks17}, BFV~\cite{bfv12}),
three public data sets (MNIST~\cite{mnist}, StanfordCars~\cite{stanfordcars}, CMUARCTIC~\cite{cmuarctic}),
and two neural networks (convolutional neural network, multilayer perceptron).
Our experiments demonstrate high effectiveness of \textsc{Chem}:
the overhead of adopting \textsc{Chem} is within a sub-second even for a 128-bit cache;
the latency reduction brought by \textsc{Chem} is 48\%--89\% for confidential inference and 67\%--87\% for encrypted model updates in federated learning.

\section{Preliminaries and Related Work}
\label{sec:prelim}

\subsection{Homomorphic Encryption}

\textit{Homomorphic encryption} (HE) is a specific type of encryption where certain operations between operands can be performed directly on the ciphertexts.
For example, if an HE scheme $he(\cdot)$ is additive,
then the plaintexts with $+$ operations can be translated into a homomorphic addition $\oplus$ on the ciphertexts.
Formally, if $a$ and $b$ are plaintexts, then the following holds:
\[
he(a) \oplus he(b) = he(a + b).
\]
As a concrete example, let $he(x) = 2^x$, and we temporarily release the security requirement of $he(\cdot)$.
In this case, $he(a+b) = 2^{a+b} = 2^a \times 2^b = he(a) \times he(b)$,
meaning that $\oplus$ is the arithmetic multiplication $\times$.

An HE scheme that supports addition is said to be \textit{additive}.
Popular additive HE schemes include Symmetria~\cite{symmetria_vldb20} and Paillier~\cite{ppail_eurocrypt99}.
The former is applied to symmetric encryption,
meaning that a single secret key is used to both encrypt and decrypt the messages.
The latter is applied to asymmetric encryption,
where a pair of public and private keys are used for encryption and decryption.
Due to the expensive arithmetical operations performed by asymmetric encryption,
Paillier is orders of magnitude slower than Symmetria.
However, Paillier is particularly useful when there is no secure channel to share the secret key among users,
which is required by symmetric encryption schemes.

An HE scheme that supports multiplication is said to be \textit{multiplicative}.
Symmetria~\cite{symmetria_vldb20} is also multiplicative using a distinct scheme than the one for addition.
Other well-known multiplicative HE schemes include RSA~\cite{rsa} and ElGamal~\cite{elgamal_tit85}.
Similarly, a multiplicative HE scheme guarantees the following equality,
\[
he(a) \otimes he(b) = he(a \times b),
\]
where $\otimes$ denotes the homomorphic multiplication over the ciphertexts.

An HE scheme that supports both addition and multiplication is called a \textit{fully HE} (FHE) scheme.
This requirement should not be confused with specific addition and multiplication parameters, such as Symmetria~\cite{symmetria_vldb20} and NTRU~\cite{ntru}.
That is, the addition and multiplication must be supported homomorphically under the same scheme $he(\cdot)$:
\[\displaystyle
\begin{cases}
    he(a) \oplus he(b) = he(a + b) \\
    he(a) \otimes he(b) = he(a \times b)
\end{cases}
\]
It turned out to be extremely hard to construct FHE schemes until Gentry~\cite{cgentry_stoc09} demonstrated such a scheme using lattice theory.
The main issue with full HE schemes is their performance;
although extensive research and development have been carried out, 
current implementations incur impractical overhead for most real-world applications.  
Popular open-source libraries of FHE schemes include HElib~\cite{helib} and SEAL~\cite{seal}.

Caching existing HE ciphertexts is also being actively researched in the database community,
e.g., Rache~\cite{otawose_sigmod23},
leveraging the binary expansion of a given plaintext in the context of outsourced databases.
However, Rache is somewhat costly due to the randomness computation being conducted repeatedly on the cached ciphertexts.
The method proposed in this work replaces the expansive computation by a polynomial number of zero ciphertexts.
HE has attracted a lot of system research interests,
such as~\cite{zhang_atc20,lphong_tifs18}.

\subsection{Radix Expansion}

In a positional numeral system, a number $x$ can be written as a summation of terms,
each of which is a product of two factors---one is the integral power of \textit{radix} $r$ and the other is the coefficient ranging from 0 to $r-1$.
Formally,
\begin{equation}\label{eq:radix}\displaystyle
x = idx_0 \cdot r^0 + idx_1 \cdot r^1 + \dots + idx_k \cdot r^k,    
\end{equation}
where $idx_i$ ($0 \le i < r$) indicates the coefficient of a specific \textit{radix entry}.
Eq.~\eqref{eq:radix} can be further expanded into an expression with only additions:
\begin{equation}\label{eq:additive}
\displaystyle
\begin{split}
    x 
    &= \underbrace{\left(r^0 + r^0 + \dots \right)}_{idx_0} 
    + \dots + \underbrace{\left(r^k + r^k + \dots\right)}_{idx_k}
    = \sum_{i=0}^k \sum_{j=1}^{idx_i} r^i,
\end{split}
\end{equation}
where $idx_i < r$. 
This purely additive form in Eq.~\eqref{eq:additive} allows us to apply additive homomorphic encryption among the radix entries rather than the original number $x$.

\subsection{Semantic Security}

When developing a new encryption scheme,
it is important to demonstrate its security level, 
ideally in a provable way.
One well-accepted paradigm with a good trade-off between efficiency and security guarantee is to assume that the adversary can launch a \textit{chosen-plaintext attack} (CPA),
meaning that the adversary can, somehow, obtain the ciphertext of an arbitrary (i.e., chosen) plaintext.
Practically speaking, however, the adversary should only be able to obtain a polynomial number of such pairs of plaintexts and ciphertexts,
assuming the adversary's machine/protocol takes polynomial time without unlimited computational resources.
Ideally, even if the adversary $\mathcal{A}$ can obtain those extra pieces of information, 
$\mathcal{A}$ should not make a \textit{distinguishably} better decision for the plaintexts than a random guess.
To quantify the degree of ``distinguishably better decision'',
\textit{negligible function} is introduced.
A function $\mu(\cdot)$ is called negligible if for all polynomials $poly(n)$ the inequality $\mu(n) < \frac{1}{poly(n)}$ holds for sufficiently large $n$'s.
The term \textit{semantic security} has a more sophisticated definition than the above but has proven to be equivalent regarding the \textit{indistinguishable} property.

For completeness, we list the following lemmas for negligible functions that will be used in later sections.
We state them without the proofs, 
which can be found in introductory cryptography or complexity theory texts.

\begin{lemma}[The summation of two negligible functions is also a negligible function]
\label{thm:neg_sum}
Let $\mu_1(n)$ and $\mu_2(n)$ be both negligible functions.
Then $\mu(n)$ is a negligible function that is defined as $\mu(n) \defeq \mu_1(n) + \mu_2(n)$.
\end{lemma}

\begin{lemma}[The quotient of a polynomial function over an exponential function is a negligible function]\label{thm:neg_quotient}

$\frac{ploy(n)}{2^n}$ is a negligible function. 
That is,
$\exists N \in \mathbb{N}^*,\; \forall n \geq N:\; \frac{ploy(n)}{2^n} < \frac{1}{poly(n)} $. 
\end{lemma}

\subsection{Machine Learning with Encrypted Tensors}

Adopting encryption, including FHE schemes, in machine learning tasks has drawn tremendous research interest.
Some notable systems include CryptoNets~\cite{ndowlin_icml16}, FHE–DiNN~\cite{fbour_crypto18}, and GAZELLE~\cite{cjuve_sec18}.
More recently, Glyph~\cite{qlou_nips20} proposed to switch between various FHE schemes to reduce the training cost of deep neural networks (DNNs).
Similarly, various works~\cite{qlou_icml21,qlou_nips20a,qlou_nips20b,qliu_nips19,rdath_pldi19,sriazi_sec19} were proposed to improve the inference performance.

While FHE represents one extreme case where the intermediate results are completely encrypted,
the so-called \textit{functional encryption}~\cite{dboneh_tcc11,sgarg_focs13} allows the neural network to decrypt the intermediate results in a specific function, such as inner products~\cite{jkatz_eurocrypt08} and more recently quadratic functions~\cite{cbaltico_crypto17}.
A semi-encrypted neural network was proposed in~\cite{tryffel_nips19} for applications such as spam filtering and privacy-preserving enforcement of content policies. 

\section{Caching Homomorphic Encryption}

\subsection{Overview}

Before describing the formal protocol,
we would like to explain the intuition behind the proposed caching method.
The first observation is that although a fully homomorphic encryption (FHE) encoding operation is usually costly,
the algebraic operation over the ciphertexts is comparatively cheaper.
With that said, if we can convert the expensive encryption operation of a given plaintext into an equivalent set of algebraic operations over existing (i.e., cached) ciphertexts,
we may obtain a performance edge.
There are two technical problems, however, in this idea.

First, which ciphertexts should we cache?
Let $he(\cdot)$ denote the FHE encoding operation.
Evidently, we can always cache only $he(1)$ and then compute $he(m)$ of $n$-bit plaintext $m$ with $\oplus_{i=1}^m he(1)$.
However, the accumulative overhead caused by a lot of homomorphic additions would at some point outweigh the cost of a single encrypting operation due to $\mathcal{O}(2^n)$ additions.
To that end, we propose to only cache a set of selective ciphertexts.
Specifically, let $r$ be a radix (and we will show how to pick $r$ soon), then the ciphertexts of $r$-power series will be pre-computed: 
$he(r^i)$, where $r^i \le 2^n$.
By doing so, the target ciphertext will be constructed through $\mathcal{O}(n)$ additions.
It should be noted that the target ciphertext at this point is merely a deterministic ciphertext,
which can be broken with a chosen-plaintext attack (CPA).
The above observation leads to the following second question.

Second, how to ensure the randomness of the ciphertext?
Randomness must be incorporated into the ciphertext to achieve a practical security level, e.g., a chosen-plaintext attack (CPA).
Informally, the randomness must be probabilistic small,
which usually takes the form of picking a polynomial number of pieces of data out of exponential parameter space.
From the above discussion, we have $n$ cached ciphertexts;
we will have to ensure the randomness comes from a set with a cardinality of something like $\mathcal{O}(2^n)$.
The key idea of our randomness construction is to cache $n$ distinct $he(0)$'s.
For each newly constructed ciphertext, we randomly add $i$ out of $n$ distinct $he(0)$'s, $0 \le i \le n$. 
Obviously, the overall space has $\sum_{i=0}^n$ $n \choose i$ or $\mathcal{O}(2^n)$ distinct $he(0)$'s.
Assuming the underlying FHE scheme is IND-CPA secure,
such a random $he(0)$ should not repeat within $2^n$ rounds of encryption.

\subsection{Protocol Description}
\label{sec:chem_protocol}

\begin{algorithm}[!t]
\SetAlgorithmName{Protocol}{}{}
\caption{Cached Homomorphic Encryption}\label{alg:rache}
\KwData{
$Ptxt[]$, $he(\cdot)$, $r$
}
\KwResult{
A tensor of ciphertexts $Ctxt[]$ such that $\forall i$, $he^{-1}\left( Ctxt[i] \right ) == Ptxt[i]$;
}
 
$m \coloneqq 2^n - 1$\;
\For{$i = 0; i <= \lfloor \log_r m \rfloor; i++$}{
    $radixes[i] \gets he(r^i)$\;
    $ct\_zeros[i] \gets he(0)$ \;
}
 
\For{i = 0; i $<$ Ptxt.size(); i++}{
    \For{$j = 0; j <= \lfloor \log_r m \rfloor; j++$}{
        $idx[j] \coloneqq (Ptxt[i]$ / $r^j)$ \% $r$\;
    }
    $Ctxt[i] \coloneqq \bigoplus_{k=0}^{\lfloor \log_r m \rfloor} \bigoplus_{j=1}^{idx[k]} radixes[k]$

    \For{j = 1; j $<$ n; j++}{
        isSwap $\xleftarrow{\$} \{0, 1\}$\;
        \If{1 == isSwap}{
            $Ctxt[i] \coloneqq Ctxt[i] \oplus ct\_zeros[j]$ \;
        }    
    }
}
\end{algorithm}

Protocol~\ref{alg:rache} formalizes the encoding procedure.
Let $n$ denote the bit-string length of plaintext messages,
i.e., $\mathcal{M} = \{0, 1\}^n$.
A tensor of plaintexts is denoted by $Ptxt[]$.
A fully homomorphic encryption scheme is denoted by $he(\cdot)$ s.t. $\forall a_i \in Ptxt[], \bigoplus_i he(a_i) = he(\sum_i a_i)$.
The radix of plaintext is denoted by $r$.
Lines 1--5 initialize the cached entries of the integral powers of radix $r$ as well as encrypted zeros for future construction of ciphertexts.
Specifically, Line 4 caches the homomorphic encryption of plaintext 0,
and will assign distinct values to $ct\_zeros$ because $he()$ is assumed IND-CPA,
meaning that encrypting the same plaintext (e.g., 0) will result in random ciphertexts.
Lines 7--10 encode each plaintext into a static ciphertext by adding up cached radix powers.
Then, the static ciphertext is randomized on Lines 11--16.

The correctness of Protocol~\ref{alg:rache} can be verified by direct computation,
as shown in Eq.~\eqref{eq:correct} in the following.
\begin{equation}
\label{eq:correct}
\begin{alignedat}{2}
Ctxt[i]
&\stackrel{(a)}{=} \left( \bigoplus_{k=0}^{\lfloor \log_r m \rfloor} \bigoplus_{j=1}^{idx[k]} radixes[k] \right) \oplus \left( \bigoplus_{k=1}^{rnd} he(0) \right) \\
& \stackrel{(b)}{=} \left( \bigoplus_{k=0}^{\lfloor \log_r m \rfloor} \bigoplus_{j=1}^{idx[k]} radixes[k] \right) \oplus he\left( \sum_{k=1}^{rnd} 0 \right) \\
&\stackrel{(c)}{=} \bigoplus_{k=0}^{\lfloor \log_r m \rfloor} \bigoplus_{j=1}^{idx[k]} radixes[k] \\
&\stackrel{(d)}{=} \bigoplus_{k=0}^{\lfloor \log_r m \rfloor} \bigoplus_{j=1}^{idx[k]} he(r^k) \\
&\stackrel{(e)}{=} \bigoplus_{k=0}^{\lfloor \log_r m \rfloor} he\left( \sum_{j=1}^{idx[k]} r^k \right) \\
&\stackrel{(f)}{=} \bigoplus_{k=0}^{\lfloor \log_r m \rfloor} he\left( idx[k] \times r^k \right) \\
&\stackrel{(g)}{=} \bigoplus_{k=0}^{\lfloor \log_r m \rfloor} he\left( (Ptxt[i] / r^k) \% r \times r^k \right) \\
&\stackrel{(h)}{=} he\left( \sum_{k=0}^{\lfloor \log_r m \rfloor} (Ptxt[i] / r^k) \% r \times r^k \right) \\
&\stackrel{(i)}{=} he\left( Ptxt[i] \right).
\end{alignedat}
\end{equation}
Recall that we denote $Ctxt[i]$ the $i$-th ciphertext,
$\bigoplus$ the homomorphic summation over a series of ciphertexts,
$\oplus$ the homomorphic summation over a pair of ciphertexts,
and $he(\cdot)$ the homomorphic encryption.
In addition, we use $rnd$ to denote the overall number of $he(0)$'s (Lines 11--16) being added to the ciphertext.
Equality $(a)$ is defined by Lines 10 and 14 in Protocol~\ref{alg:rache}.
Equal signs $(b)$, $(c)$, $(e)$ , and $(i)$ are due to the definition of fully homomorphic encryption.
Equality $(d)$ is due to Line 3 of Protocol~\ref{alg:rache}.
Equality $(f)$ is because variable $j$ does not show up in the term $r^k$.
Equality $(g)$ is due to Line 8 of Protocol~\ref{alg:rache}.
Equality $(i)$ is due to the definition of radix expansion.

The time complexity of Protocol~\ref{alg:rache} is as follows.
Lines 1 -- 5 take $\mathcal{O}(n)$.
Similarly, both Lines 7--8 and Lines 12--16 take $\mathcal{O}(n)$.
Line 6 implies $\mathcal{O}(n)$ iterations,
which means Lines 6--17 take $\mathcal{O}(n^2)$.
As a result, the overall time complexity is $\mathcal{O}(n^2)$.
It should be noted that the above analysis assumes the homomorphic encryption takes a constant time $\mathcal{O}(1)$;
in practice, an $he()$ operation takes significantly more time than regular arithmetic operations.

\subsection{Security Analysis}
\label{sec:chem_security}

\subsubsection{Threat Model}
\label{sec:chem_security_threat}

We assume the adversary can obtain up to $ploy(n)$ pairs of $(plaintext, ciphertext)$.
The plaintext could be arbitrarily chosen by the adversary.
That is, the adversary can launch a \textit{chosen-plaintext attack} (CPA).
For example, the attacker might be able to retrieve a copy of the cached radix powers through a memory attack or social engineering.
However, we do assume the security keys used in the underlying FHE scheme are appropriately updated to avoid the \textit{key fatigue} issue.

We assume the adversary can only run probabilistic polynomial-time (PPT) algorithms.
We also assume the primitive FHE scheme is IND-CPA secure,
such as Paillier~\cite{ppail_eurocrypt99} and Symmetria~\cite{symmetria_vldb20}.
In particular, the encryption function of the FHE must be probabilistic---a requirement for a scheme to be IND-CPA secure.

\subsubsection{Provable Security}
\label{sec:chem_security_proof}

This subsection demonstrates the semantic security of \textsc{Chem} (Protocol~\ref{alg:rache}).
We first explain the intuition why \textsc{Chem} is IND-CPA secure and then sketch the proof.

Recall that \textsc{Chem} pre-computes and caches $\log_r 2^n$ radix entries.
If we assume the system picks $r=2$ 
(later we will show that $r=2$ is optimal in the worst case),
then the scheme will simply cache $n$ radix-power entries.
Therefore, those ciphertexts cached by \textsc{Chem} should not significantly help the adversary---who presumably runs a probabilistic polynomial-time (PPT) Turing machine---as the overall space is exponential, $2^n$ by the assumption of the $n$-bit security parameter in the primitive FHE scheme.

Technically, we want to \textit{reduce} the problem of breaking an original FHE scheme $\Pi$ to the problem of breaking the \textsc{Chem} counterpart $\widetilde{\Pi}$.
In other words, if a PPT adversary $\mathcal{A}$ takes an algorithm $alg$ to break $\widetilde{\Pi}$, 
then $\mathcal{A}$ can efficiently (i.e., in polynomial time) construct another algorithm $alg'$ that calls $alg$ as a subroutine to break $\Pi$ as well (simulating $alg'$ with $alg$).
However, if we already know that the original FHE is IND-CPA secure,
the above cannot happen---leading to a contradiction that proves the IND-CPA secure of \textsc{Chem}.

We formalize the above reasoning in the following proposition.
\begin{proposition}\label{thm:rache}
If fully homomorphic encryption $\Pi$ is semantically secure under the threat model of chosen-plaintext attack (IND-CPA),
then its corresponding \textsc{Chem}-extension $\widetilde{\Pi}$ defined in~Protocol~\ref{alg:rache} is also IND-CPA secure.
\end{proposition}

\begin{proof}[Proof (sketch)]
Notionally, let $CPA^{\mathcal{A}}_{X}$ denote the indistinguishability experiment with scheme $X$.
The probability for $\mathcal{A}$ to successfully break $\Pi$ and $\widetilde{\Pi}$ are $Pr\left[CPA_{\Pi}^\mathcal{A} = 1 \right]$ and $Pr\left[CPA_{\widetilde{\Pi}}^\mathcal{A} = 1 \right]$, respectively.
By assumption, the following inequality holds:
\begin{equation}\label{eq:cpa_b}
Pr\left[CPA_{\Pi}^\mathcal{A} = 1 \right] \leq \frac{1}{2} + \epsilon,     
\end{equation}
where $\epsilon$ is a negligible probability.
By comparing $\Pi$ and $\widetilde{\Pi}$,
the latter yields $2n$ additional pairs of plaintexts and ciphertexts out of the total $2^n$ possible pairs:
$n$ for $r$-powers and another $n$ for zeros.
Therefore, the following inequality holds:
\begin{equation}\label{eq:cpa_diff}
    Pr\left[CPA_{\widetilde{\Pi}}^\mathcal{A} = 1\right] - Pr\left[CPA_{\Pi}^\mathcal{A} = 1\right] \leq \frac{poly(n)}{2^n}.
\end{equation}
Combining Eq.~\eqref{eq:cpa_b} and Eq.~\eqref{eq:cpa_diff}
yields the following inequality:
\[
Pr\left[CPA_{\widetilde{\Pi}}^\mathcal{A} = 1\right] \leq \frac{1}{2} + \epsilon + \frac{poly(n)}{2^n}.
\]
Now, we only need to show that the summation of the last two terms, $\epsilon + \frac{poly(n)}{2^n}$, is negligible.
According to Lemmas~\ref{thm:neg_sum} and~\ref{thm:neg_quotient}, 
this is indeed the case.
Therefore, the probability for the adversary $\mathcal{A}$ to succeed in the $CPA_{\widetilde{\Pi}}^\mathcal{A}$ experiment is only negligibly higher than $\frac{1}{2}$,
proving the semantic security of \textsc{Chem}, as claimed.
\end{proof}

\subsection{Parametrization}
\label{sec:chem_para}

So far, we have been using the variable $r$ to denote the underlying radix.
We are interested in having some estimation of $r$ regarding practical performance.
This section will investigate an optimal radix in the worst case.

Let $m \ge 2$ denote the maximal number to be encrypted in the machine learning task.
Let $r \ge 2$ denote the radix or base of the homomorphic encryption.
Obviously, given an arbitrary number $x$, where $0 \le x \le m$,
there are $k + 1$ radix entries:
$r^0$, $r^1$, $\dots$, $r^k$,
where $k = \lfloor \log_r^m \rfloor$.
Let $0 \le \kappa \le k$.
In the worst case, each $r^{\kappa}$ radix-entry incurs $r - 2$ times of homomorphic addition,
i.e., when computing $(r - 1) \cdot x^{\kappa}$.
Since one more homomorphic addition needs to be taken for the summation of each radix,
the overall times of homomorphic addition, in the worst case when $m$ is one less than the next integral power of $r$ (i.e., $\lfloor \log_r^m \rfloor = \log_r^{m+1} - 1$), is 
\begin{alignat*}{2}
f(r)&= (r-2)(k+1) + k 
        = (r-1)k + r - 2\\
    &= (r-1)(\log_r^{m+1} - 1) + r - 2   = (r-1)\log_r^{m+1} - 1.
\end{alignat*}

We will find out the optimal $r$ that minimizes $f(r)$.
We take the first-order derivative of $f(r)$ in the following: 

\begin{alignat*}{2}
f'(r) 
&= \frac{d}{dr} f(r) \\
&= \frac{d}{dr} \left( (r - 1)\log_r^{m+1} - 1 \right)\\
&= \frac{d(r-1)}{dr}\cdot \log_r^{m+1} + (r-1)\cdot \frac{d}{dr} \left( \log_r^{m+1} \right)\\
&= 1\cdot \log_r^{m+1} + (r-1)\cdot \frac{d}{dr} \left(\frac{\ln (m+1)}{\ln r} \right) \\
&= \log_r^{m+1} + (r-1)\ln (m+1) \cdot \frac{d}{dr} \left( \left(\ln r \right)^{-1} \right) \\
&= \log_x^{m+1} - (r-1)\ln (m+1) \cdot (\ln r)^{-2} \cdot \frac{1}{r} \\
&= \ln (m+1) \cdot (\ln r)^{-2} \cdot r^{-1} \cdot \left( r\ln r - r + 1 \right).
\end{alignat*}
The stationary point is therefore the solution to $f'(r) = 0$, or
\[\displaystyle
g(r) = r\ln r - r + 1 = 0,
\]
which yields $r = 1$.
Since we require $r \ge 2$, we need to find another qualified radix.
First, we calculate $g(2)$:
\[\displaystyle
g(2) = 2\ln 2 - 2 + 1 \ge 2\times 0.69 - 1 > 0.
\]
Then, let $r \ge 3$, therefore $\ln r > 1$, which yields:
\[\displaystyle
\left.g(r)\right\rvert_{r \ge 3} = r\ln r - r + 1 = r(\ln r - 1) + 1 > 0.
\]
Note that by definition, the following equation holds:
\[\displaystyle
f'(r) = \ln (m+1) \cdot (\ln r)^{-2} \cdot r^{-1} \cdot g(r).
\]
If we assume $m \ge 2$, then $\ln (m+1) > 0$.
Both $(\ln r)^{-2}$ and $r^{-1}$ factors are obviously positive.
Therefore, $f'(r)$ is always positive, meaning that $f(r)$ is a monotonically increasing function.
It follows that radix $r = 2$ leads to the minimum number of homomorphic additions in the worst case.

\section{Evaluation}
\label{sec:eval}

\subsection{Implementation}

We implement \textsc{Chem} with OpenMined TenSEAL~\cite{tenseal2021},
a tensor encryption library for machine learning.
While TenSEAL is implemented with C++ and Python,
as its name suggests, TenSEAL is enlightened by Microsoft SEAL~\cite{seal}.
At the writing of this paper, TenSEAL supports two types of fully homomorphic encryption (FHE) schemes: CKKS~\cite{ckks17} and BFV~\cite{bfv12}.
While both CKKS and BFV are based on learning with errors (LWE) in lattice cryptography (e.g., polynomial rings),
CKKS is optimized for float numbers and BFV was originally designed for integers.
Indeed, CKKS is slightly slower than BFV due to the additional support for float numbers.
However, the support and optimization for float numbers are usually highly desired features.

\begin{table*}[t]
  \caption{Datasets for evaluation}
  \label{tbl:dataset}
  \centering
  \begin{threeparttable}
  \begin{tabular}{lllll}
    \toprule
    \multicolumn{2}{c}{Dataset}                   \\
    \cmidrule(r){1-2}
    Name     & Description     & Size (test set) & Shape & Nonempty Rate\\
    \midrule
    MNIST~\cite{mnist}   & Handwriting digits  & 10,000  & [28$\times$28] & 17.90\%     \\
    StanfordCars~\cite{stanfordcars}     & Car images & 8,041 & [360$\times$640]$^*$ & 98.73\%     \\
    CMUARCTIC~\cite{cmuarctic}     & English speech & 1,132 & [64,321]$^*$ & 99.72\%\\
    \bottomrule
  \end{tabular}
  \begin{tablenotes}[para,flushleft]
  $^*$The shapes vary for different data samples, although they are ``close'' to each other.
  \end{tablenotes}
  \end{threeparttable}  
\end{table*}

\newcommand{\figwidth}{0.49}
\begin{figure*}[!t]
\begin{minipage}[t]{\figwidth\linewidth}
	\includegraphics[width=\linewidth]{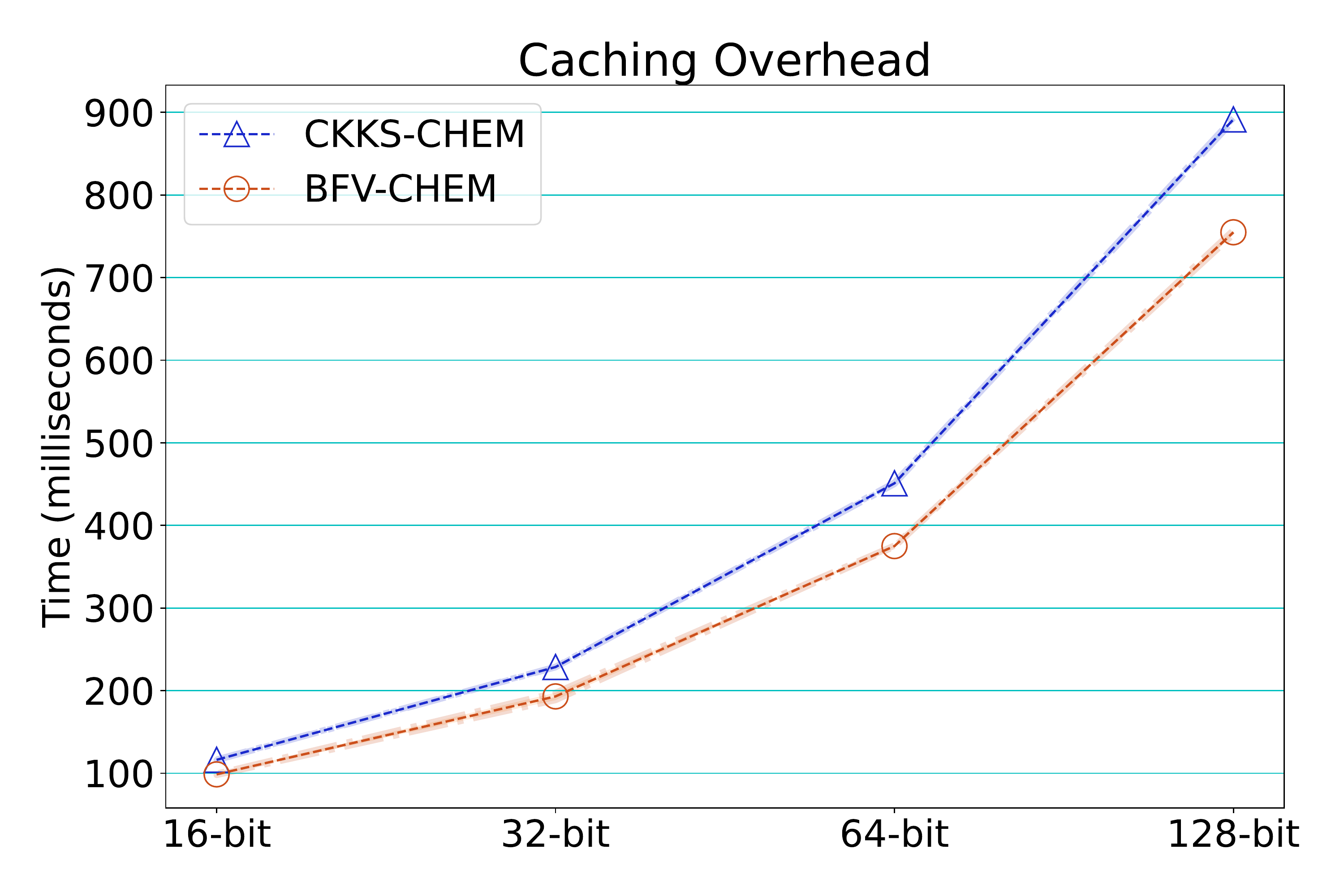}
	\caption{CKKS- and BFV-\textsc{Chem} overhead on different cache sizes.}
	\label{fig:overhead}
\end{minipage}%
    \hfill%
\begin{minipage}[t]{\figwidth\linewidth}
	\includegraphics[width=\linewidth]{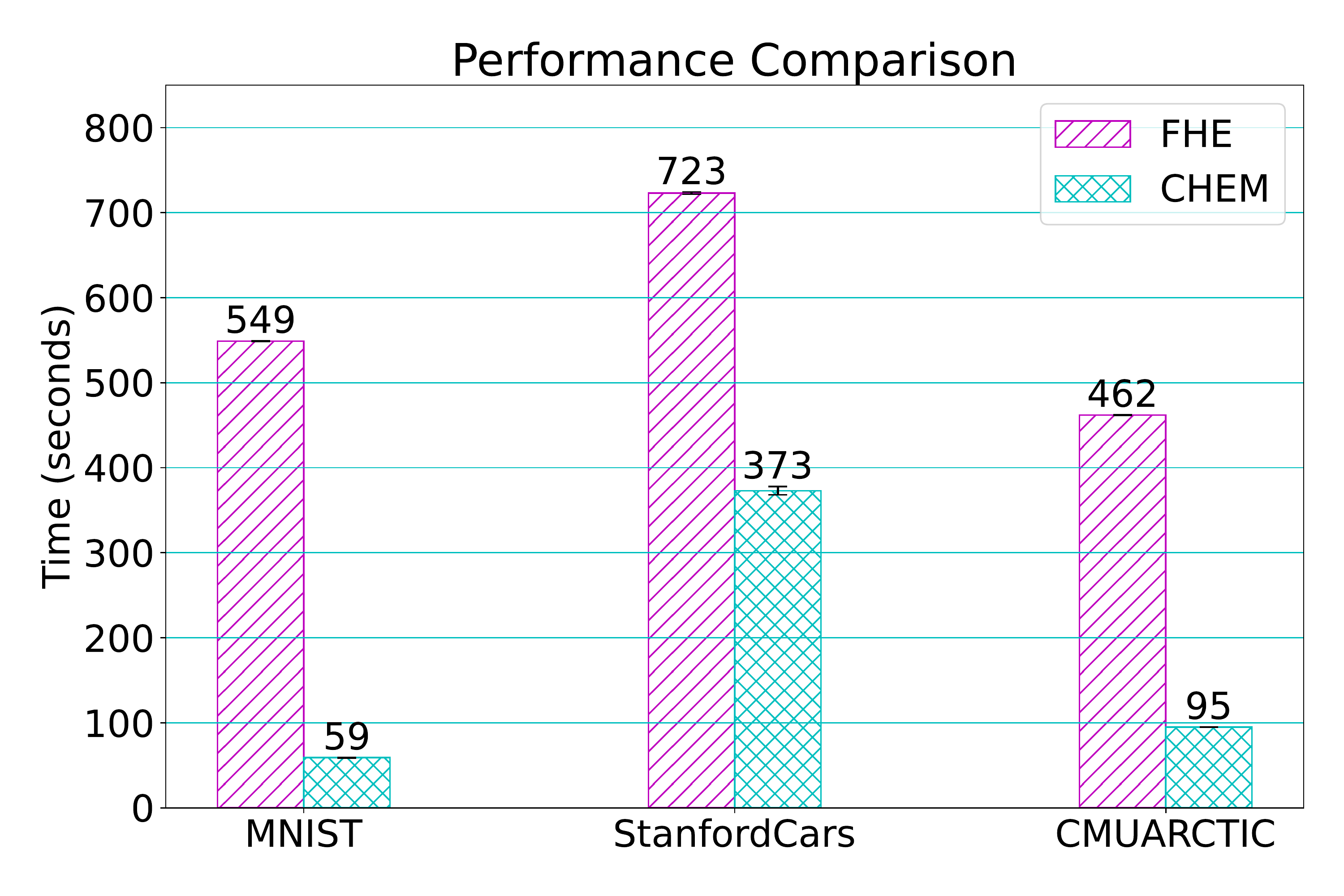}
	\caption{Performance comparison between FHE and \textsc{Chem} on different data sets.}
	\label{fig:performance}
\end{minipage} 
\end{figure*}

The proposed \textsc{Chem} can be applied to an \textit{arbitrary} FHE scheme as long as the plaintext can be converted into a bitstring.
We have implemented \textsc{Chem} for both CKKS and BFV.
Thanks to the Python interface realized in TenSEAL,
the core implementation of \textsc{Chem} consists of only about 1,500 lines of Python code on top of a recently published FL framework MPI-FL~\cite{lwang_sc22} for high-performance computing clusters.

\subsection{Experimental Setup}

Experiments are carried out on CloudLab~\cite{cloudlab}.
We use the \texttt{m510} instance,
which is equipped with 8-core Intel Xeon D-1548 CPUs, 
64GB ECC memory, 
and 256GB NVMe flash storage.
The operating system image is Ubuntu 20.04.4.
Important libraries installed in the system include:
Python~3.8.10,
PyTorch~1.11.0,
Numpy~1.22.3, 
and
Scipy~1.8.0.
Our C++ compiler is GNU g++~9.4.0.

\begin{table*}[htbp]
  \caption{Neural network models for evaluation}
  \label{tbl:model}
  \centering
  \begin{threeparttable}
  \begin{tabular}{llll}
    \toprule
    Model & Optimizer & Learning Rate & Topology and Function \\
    \midrule
    CNN & SGD & 0.01 & conv: $1 \rightarrow 10 \rightarrow 20$, kernel $5 \times 5$; fc: $320 \rightarrow 50 \rightarrow 10$      \\
    MLP & SGD & 0.01 & $784 \rightarrow 64 \rightarrow 10$, ReLU, Softmax     \\
    \bottomrule
  \end{tabular}
  \end{threeparttable}  
\end{table*}

\begin{figure*}[!t]
\begin{minipage}[t]{\figwidth\linewidth}
	\includegraphics[width=\linewidth]{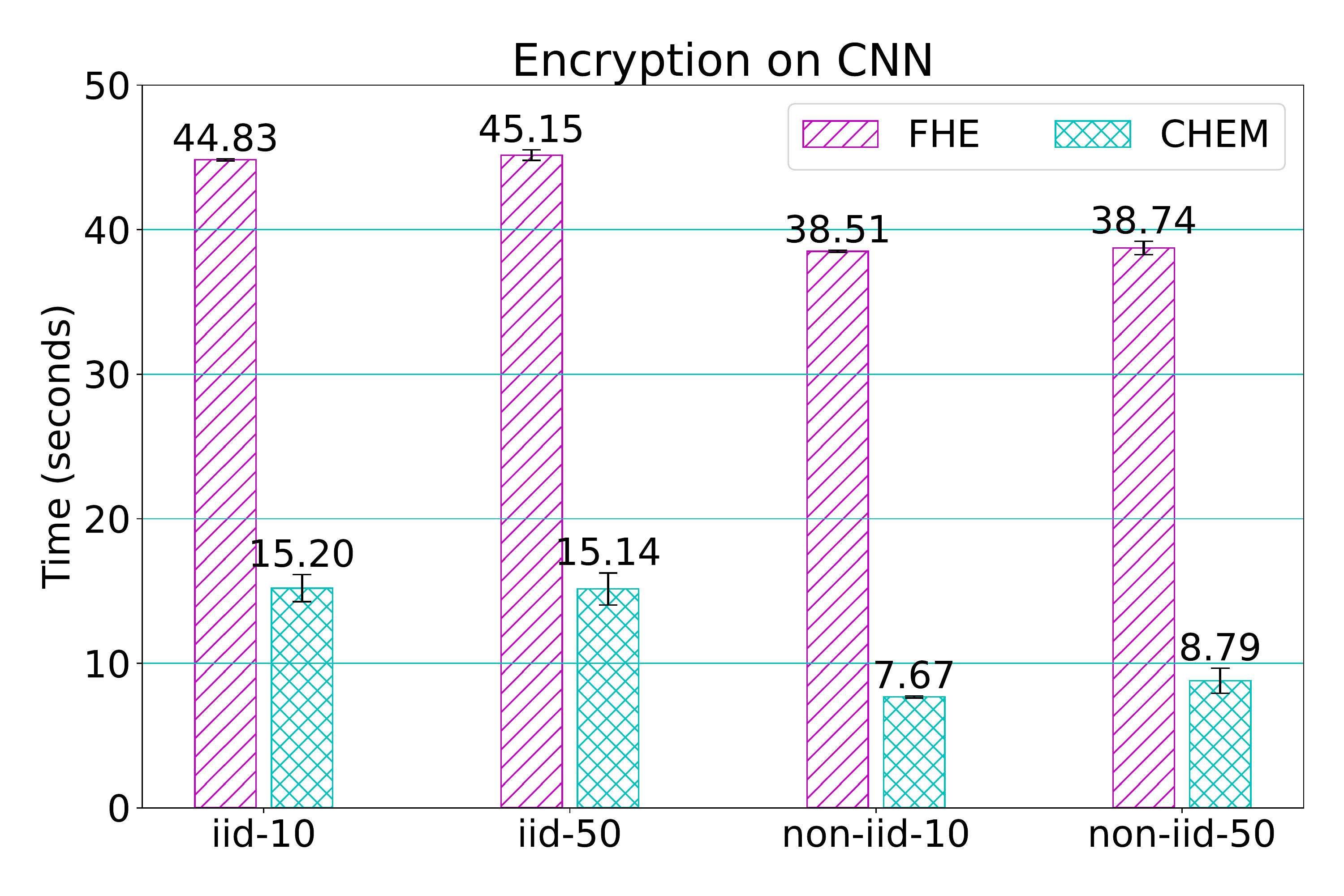}
	\caption{Performance comparison between FHE and \textsc{Chem} in federated convolutional neural networks (CNNs).}
	\label{fig:fl_cnn}
\end{minipage}%
    \hfill%
\begin{minipage}[t]{\figwidth\linewidth}
	\includegraphics[width=\linewidth]{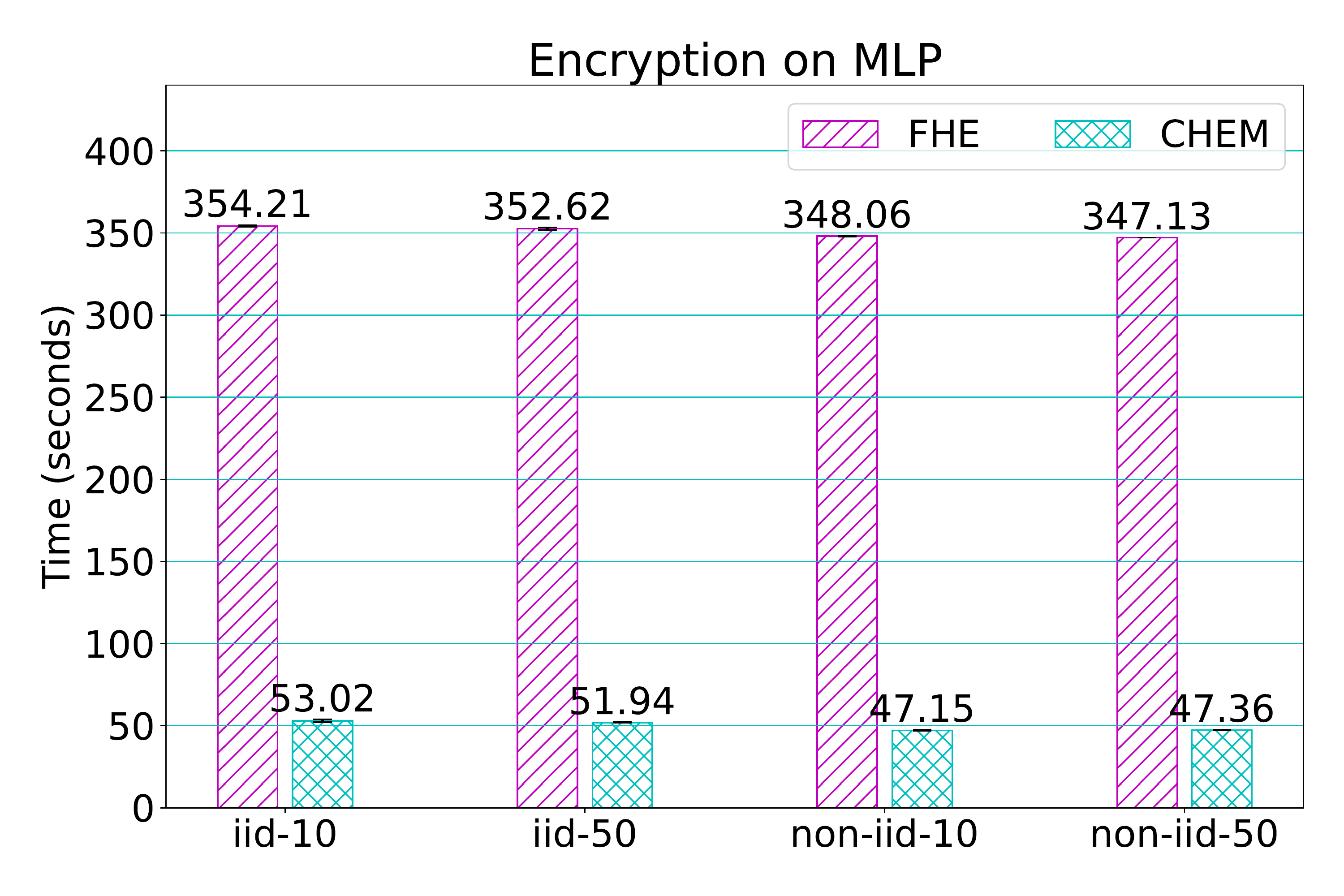}
	\caption{Performance comparison between FHE and \textsc{Chem} in federated multilayer perceptron (MLP) networks.}
	\label{fig:fl_mlp}
\end{minipage} 
\end{figure*}

Our experiments take three data sets all from the official PyTorch repository: 
MNIST~\cite{mnist}, StanfordCars~\cite{stanfordcars}, and CMUARCTIC~\cite{cmuarctic}.
We summarize the key properties of these data sets in Table~\ref{tbl:dataset}.
It should be noted that only the MNIST data set maintains a fixed shape of the tensors;
for those variable-shape data sets, 
we list the shape of a random data sample.
While other properties are self-explanatory, 
the \texttt{Nonempty rate} column shows the rate of nonempty values (i.e., non-zero) for the data samples.
Two baseline FHE schemes in our evaluation are BFV~\cite{bfv12} and CKKS~\cite{ckks17}.
We repeat all experiments at least three times and report the average along with the standard derivation (which might be unnoticeable in the plots due to the small variance).
By default, CKKS is used as the baseline FHE scheme unless otherwise stated.

\subsection{Caching Overhead}

We start our evaluation by quantifying the performance overhead for pre-computing the radix entries.
Figure~\ref{fig:overhead} illustrates the overhead for caching different numbers of radix entries ranging from 16-bit to 128-bit.
Indeed, the overhead for both CKKS- and BFV-based \textsc{Chem} caching is proportional to cache size because our implementation does not currently employ any parallel-processing library (e.g., OpenMP~\cite{openmp}).
The gap between the CKKS-\textsc{Chem} and the BFV-\textsc{Chem} is also expected:
this is due to the more sophisticated computation involved in CKKS for handling float numbers.

Figure~\ref{fig:overhead} also shows that the overhead is within a sub-second even for a 128-bit cache.
Note that 128-bit is considered a reasonably secure parameter for modern cryptosystems.
As we will see shortly in the next section,
the performance gained by adopting \textsc{Chem} could turn out to be orders of magnitude higher than the overhead.

\subsection{Encrypted Inference}
\label{sec:eval_performance}

Figure~\ref{fig:performance} evidently demonstrates the effectiveness of \textsc{Chem}:
encoding MNIST with \textsc{Chem} only takes about 11\% of the time needed for the original FHE scheme;
for StanfordCars, \textsc{Chem}-based encryption takes about 52\% of the FHE time;
and the ratio becomes 21\% for the CMUARCTIC data set.
That is, the time saving is 89\%, 48\%, and 79\%, 
respectively, for MNIST, StanfordCars, and CMUARCTIC.

There are two main reasons why the improvements for different data sets exhibit discrepancies.
First, recall that the majority cost of \textsc{Chem} stems from the homomorphic addition over cached radix-powers.
This implies that the overhead largely depends on how much the plaintext aligns away from those radix powers.
Indeed, this alignment is application-specific.
Second, as a special case of the first reason, 
the caching effectiveness is also highly sensitive to the zeros in the plaintext.
If the plaintext comprises a lot of zeros,
Lines 7--10 can be skipped in Protocol~\ref{alg:rache},
which implies faster encryption.
Combining both reasons, we now have a better sense of the fact that MNIST benefits the most from \textsc{Chem} because the non-empty rate of MNIST is only 17.9\%,
much lower than the other two data sets (98-99\%).
As for the difference between StanfordCats and CMUARCTIC,
it can be best explained by the alignment discrepancy,
i.e., the first reason discussed above.

\subsection{Encrypted Models in Federated Learning}

Last but not least, we report the effectiveness of \textsc{Chem} in the context of federated learning.
We will use two neural network models for evaluating \textsc{Chem} and FHE in federated learning:
a convolutional neural network (CNN) and a multilayer perceptron (MLP).
We will focus on the MNIST data set in federated learning experiments.
Table~\ref{tbl:model} lists some of the properties of both networks.

We fix the overall number of clients (i.e., the number of local models being concurrently trained) to 30.
We vary the fraction of users being involved in a single round of model updates between 10\% and 50\%.
We also distribute the training data into either independent and identically distributed (iid) partitions or non-iid partitions.
A non-iid partition can be achieved by sorting the labels of data and then dispersing them of the same or adjacent labels to the same local model.
We apply FHE and \textsc{Chem} to the updated weights of two convolution layers in CNN and both the input and hidden layers in MLP.

Figures~\ref{fig:fl_cnn} and~\ref{fig:fl_mlp} report the latency of FHE and \textsc{Chem} for a single round of aggregation among the local models in federated learning.
The $x$-label indicates the $(dataDistribution, userFraction)$ pair;
for example, `\texttt{iid-50}' means that data samples are distributed independently and identically into 50\% clients;
that is, a synchronization round involves $30 \times 50\% = 15$ local modes. 
Figures~\ref{fig:fl_cnn} show that \textsc{Chem} reduces the encryption cost of FHE by 67\%--81\%.
Similarly for MLP, 
Figures~\ref{fig:fl_mlp} show that \textsc{Chem} reduces the encryption cost of FHE by 84\%--87\%.
For both models, the fractions of users do not make a significant impact on the performance.

\section{Conclusion and Future Work}

This paper proposes the \textsc{Chem} protocol to cache tensor ciphertexts in machine learning tasks such that the latter can be constructed from a pool of ciphertexts rather than touching on expensive encryption operations.
The IND-CPA security of \textsc{Chem} is proven,
and an optimal radix is analyzed.
Experimental results show that adopting \textsc{Chem} only incurs a sub-second overhead and yet reduces the encryption latency by 48\%--89\% for confidential inference and 67\%--87\% for encoding local models in federated learning, respectively.

Our future work will focus on exploring more encryption schemes customized for machine learning tasks,
especially for federated learning systems.
Another important and yet not implemented in \textsc{Chem} is the parallel processing of additive homomorphic encryption over the ciphertexts. 

\newpage
\bibliographystyle{named}
\bibliography{ref_new}

\end{document}